\documentclass[12pt,a4paper]{article}
\usepackage{graphicx}
\usepackage[T1]{fontenc}
\usepackage{amssymb}
\usepackage{inputenc}
\usepackage{url}
\usepackage{array}
\usepackage{graphicx}
\usepackage{pstricks}
\usepackage{pst-node}
\usepackage{amssymb}
\usepackage{amsmath}
\usepackage{eurosym}
\usepackage{geometry}
\newgeometry{tmargin=3cm, bmargin=3cm, lmargin=1.5cm, rmargin=1.5cm}
\newtheorem{df}{Definition}

\begin{document}
\title{DO CLASSICAL (OR QUANTUM) TRANSITIVE PREFERENCES ALWAYS RESULT IN INDIFFERENT DIVISIONS?}

\author{Marcin Makowski $^{1,}$*, Edward W. Piotrowski $^{1}$** and Jan S\l{}adkowski $^{2}$\\
 \small 1. Institute of Mathematics, University of Bia\l ystok,\\\small Akademicka 2, PL-15424, Bia{\l}ystok, Poland
\\\small $*$ makowski.m@gmail.com, $**$ qmgames@gmail.com
\\ \small 2. Institute of Physics, University of Silesia,\\\small Uniwersytecka 4, 40-007 Katowice, Poland\\
\small jan.sladkowski@us.edu.pl}
\maketitle

\begin{abstract} The transitivity of preferences is one of the basic assumptions used in the theory of games and decisions. It is often equated with rationality of choice and is considered useful in building rankings. Intransitive preferences are considered paradoxical and undesirable. This problem is discussed by many social and natural sciences. The paper discusses a simple model of sequential game in which two players in each iteration of the game choose one of the two elements. They make their decisions in different contexts defined by the rules of the game. It appears that the optimal strategy of one of the players can only be intransitive! (the so-called \textsl{relevant intransitive strategies}.) On the other hand, the optimal strategy for the second player can be either transitive or intransitive. A quantum model of the game using pure one-qubit strategies is considered. In this model, an increase in importance of intransitive strategies is observed -- there is a certain course of the game where intransitive strategies are the only optimal strategies for both players. The study of decision-making models using quantum information theory tools may shed some new light on the understanding of mechanisms that drive the formation of types of preferences.\end{abstract}





\section{Introduction}

Games have always fascinated scholars, often contributing to the development of new theories \cite{ecbi}. In fact, the attempts to construct a systematic theory of rational behaviour are focused on games as simple examples of human rationality. The attractiveness of such approach to the analysis the interactions between rational players (a problem reflected in many fields of science) reveals itself in its various applications \cite{stt}. Game theory methods have been used in areas such as military science, biology, economics and other social sciences. Since the very beginning, the game theory has been closely connected with the information theory \cite{inf}. Therefore, during the development of the theory of quantum information \cite{rq}, a quantum game theory has emerged in a natural way. In its general form, pure strategies are identified with Hilbert space vectors (pure states) --- qubits or qubit systems. Mixed strategies are represented by convex combinations of pure states. Replacing classical probabilities used in the game theory with quantum probability amplitudes provides many interesting opportunities arising from superposition and entanglement. The idea of constructing quantum models of games is intensely developing nowdays. Numerous examples show the rapid rate with which the situation of players changes after obtaining access to quantum technology and how the fundamental limitations of classical models can be overcomed thanks to the properties of quantum processes \cite{re4,rd5,rA6,rP7,r8}. 

This paper considers a simple model of a repeated game in which, at each stage, both players divide between themselves a set consisting of three goods. With reference to the earlier results \cite{r1,r2,r2n,En,El,div} influenced by the remarks of Hugo Steinhaus \cite{ST}, the players are referred to as 'cats' and goods are referred to as 'foods'. The main goal of our analysis is to examine how rules of the game affect the type of optimal strategies for the players (whether they are transitive or intransitive). The differences arising from the adoption of a different way of sourcing strategies are also examined. One of them is a mixed strategy based on one bit (classical variant) and the other is one-qubit pure state (quantum variant).

Before we proceed to the formal description of the game, let us highlight the main ideas of transitivity --- intransitivity. This is done in the next subsection. The second subsection of this introduction is devoted to review of recent results that have led us to formulate  the main results.
  
\subsection{Intransitivity}

Any relation $\succ$ between the elements of a certain set is called transitive if  
\begin{equation}\label{relacja} 
 A\succ B \wedge B\succ C \Rightarrow A\succ C. 
\end{equation}
\noindent
is fulfilled for any three elements $A$, $B$, $C$. If this condition is not fulfilled then the relation will be called
intransitive (not transitive).

The problem of transitivity (intransitivity)  stems from various fields of research. 
There is an opinion that people who make decisions relying on rational reasoning, should make decisions in determined and linear order   \cite{trra,wr}. Transitivity of preferences indicates the way of choosing according to the ``logical order''. There is also a hypothesis that many animals (also people) follow transitive inference rules (choosing \textit{A} over \textit{C} on the basis of knowing that \textit{A} is better than \textit{B} and \textit{B} is better than \textit{C}). This type of reasoning has been confirmed in several animal species \cite{r33}.

One of the main arguments against intransitiveness is the so-called ``money pump'' \cite{tull}. On the other hand, some modification of utility theory and decision theory which dispense of the transitivity assumption have been considered  \cite{r23, tw}.

The beginnings of research on intransitive orders probably dates back to 1785 when Jean Condorcet published his work: \textit{Essay on the Application of Analysis to the Probability of Majority Decisions} (1785), in which he analyzed the paradox of voting. He concluded that collective preferences can be intransitive even if the preferences of individual voters are not. Analysis of this paradox led Kenneth Arrow (Nobel Prize winner in economy) to prove that an election procedure
which would perfectly agree with basic postulates of democracy does not exist \cite{r24}.

In psychology, which also attempts to explain, inter alia, the decision making
process, the special interest is focused on a broadly understood relation
of superiority or domination (preference) \cite{lek, T}. Does the fact that A dominates
over B and B dominates over C, imply that A dominates over C? It turns out that the answer is not obvious and depends on a particular situation. The discussion on the concept of dominance has also continued using the tools provided by exact sciences with probabilistic models as good examples, e.g.~Efron's Dice \cite{Gar}. The game proposed by Walter Penney \cite{r35} is another example of intransitivity in probabilistic models. Intransitivity may explain the processes occurring in Nature. Rivalry between species may be intransitive. For example, in the case of fungi, Phallus impudicus replaced
Megacollybia platyphylla, M.~platyphylla replaced Psathyrella hydrophilum, but P.~hydrophilum replaced P.~impudicus \cite{r15}. Similarly, we can explain the stability of the population of lizards \cite{jasz} or experiments with bees which make intransitive
choices between flowers \cite{r16}.
Intransitivity models appear also  in many seemingly distant sciences including philosophy \cite{trra}, operations research \cite{ope},  thermodynamics \cite{Ki}, quantum theory \cite{r2n}, logic \cite{H}.

Research on the rational decision-making process is complex and consists of many various concepts. There is a large literature with the discussions on rational choice \cite{Trv}. This is a difficult problem because it turns out that the assumptions made by theoretician (such as completeness, transitivity, independence of irrelevant alternatives or others) are often broken in the course of experiments with humans.  Thus, people have constructed theories of choice without transitivity or other assumptions. 
In this paper we focus on one of the most important issues in the theory of decision --- the transitivity/intransitivity of the choice. Reflections on this occurs in the context of studies of various types of preferences, state-dependent preferences \cite{St}, context-dependent preferences \cite{trS}.

Intransitive issue is still not well understood. It is worth to study this using tools of quantum game theory. This new look can lead to many interesting conclusions that can be used in research on decision-making process simulation and other biological mechanisms.

\subsection{Earlier results. ``I cut, you choose'' game.}

In his diary, Hugo Steinhaus mentions Pitts experiments with cats. It turns out that that the cat, facing
the choice between fish, meat and milk prefers fish to meat, meat to milk, and milk to fish! Steinhaus thought that the cat provided itself with a balanced diet, thanks to the above-mentioned food preferences.  This is one of the key factors needed to maintain good health.
Hence, in our model of the game, the players are cats choosing among foods. Obviously, this is only an illustration of the problem which can be interpreted in different ways (e.g. in relation to other goods, or as an electoral issue).

In the paper \cite{r1} was considered a classic model of a game in which the player (cat) are offered three types of foods,  every time in pairs of two types. Optimal strategy was defined as one that leads to a balanced diet (equal distribution of the frequency of the occurrence of a particular food in cat's diet). The offering player (Nature) was not interested in the result of the game. 
The quantitative analysis of various types (intransitiv or transitive) optimal strategies indicated the advantage of the transitive strategies. Intransitive strategies represent a significant part of all optimal strategies, but in the situation that favors optimal strategies we can always find the strategy that determines the transitive order (with identical result, under the same conditions).
This situation changed in the quantum variant of the game \cite{r2}, which reveals the existence of the so-called \textit{relevant intransitive strategies}, defined as follows \cite{El}:

\begin{df}
The intransitive strategy will be called the relevant strategy,
if there is no transitive strategy of the same consequences with the same
assumption.
\end{df}
In the paper \cite{El} was considered electoral interpretation of the game (see also \cite{kier}). The decrease of importance of intransitive orders which accompanies the growth of support for one of the candidates \footnote{If voters start to clearly prefer one of the candidates.}  turned out to be an interesting property of the quantum game model. The use of an entangled state to the model construction causes an increase in the importance of transitive strategies.

\begin{figure}[h]
\includegraphics[
          height=4.5in,
         width=5.9in]{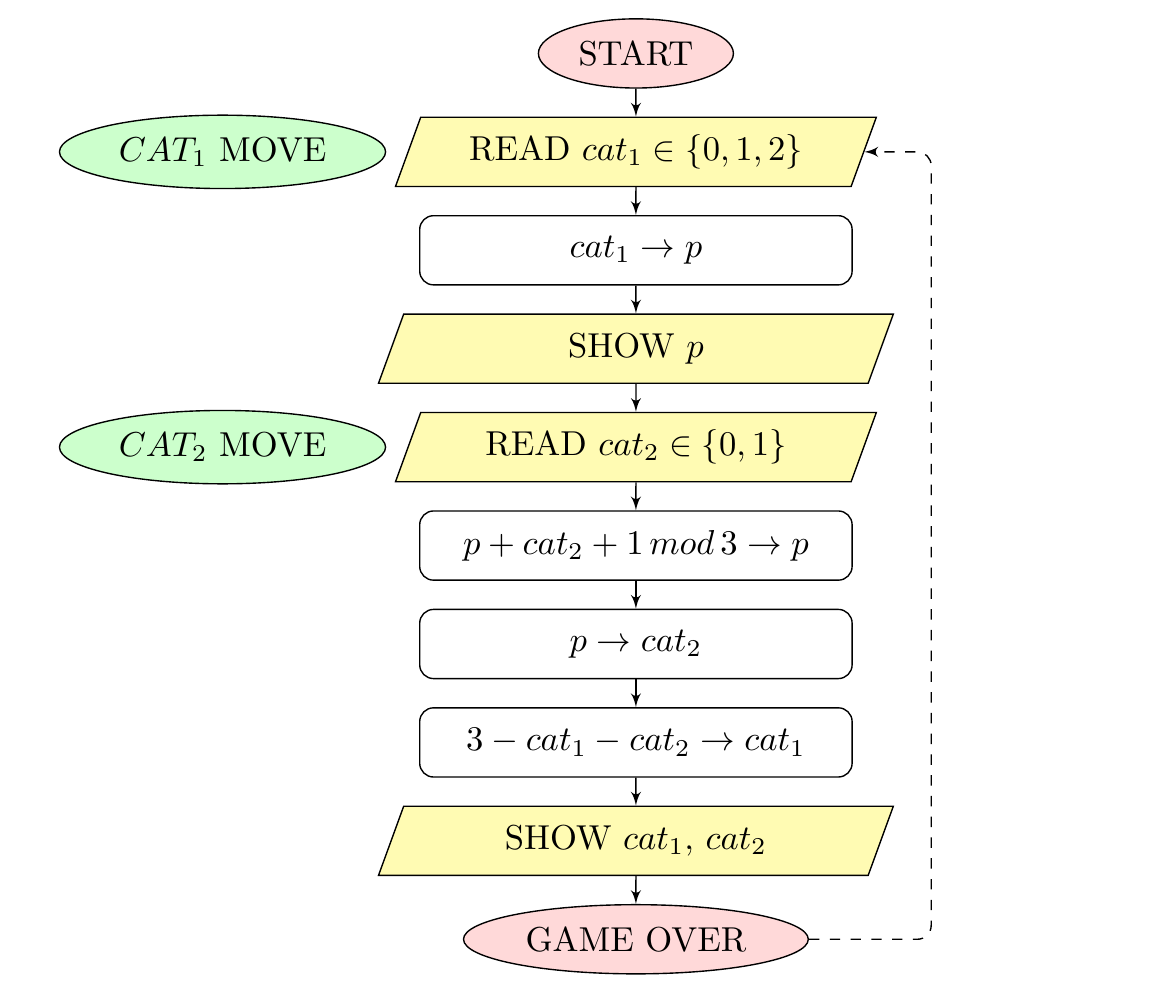}
			\caption{Flow diagram of each iteration of the 'I cut, you choose' game.}\label{ggwo}
\end{figure}

It turns out that  intransitive relevant strategies are not only characteristic of quantum models. In the paper \cite{div} offering player  (Nature) has been replaced by a rational player. In this game both players (Cat 1 and Cat 2) divide the set of three foods according to the ''I cut, you choose'' method \cite{Brams, ST1} (see Fig.\ref{ggwo}). The first player, chooses and rejects one
of the foods. Then the second player,  selects and consumes one of the remaining two foods. The first player eats the food that is left. The optimum consists (as in previous models) in not distinguishing any of the three foods. Each of the foods is equally important to each of the players.

In this variant of the game, only the second player (Cat 2) makes choices between the pairs of foods offered to him. Therefore, transitive and intransitive strategies can be analysed only in case of this player. It appears that his optimal strategies are only intransitive strategies (\textsl{relevant intransitive strategies}.). 

In this article, a modification of the above game is analysed. It consists in letting the player choose the food selected at the first stage of the game, as well as the food rejected by the other player. This modification allows both players to determine their preferences with regard to the pairs of foods. It will enable exploring the types of strategies and their availability for individual players.

\section{Classical model}

\begin{figure}[h]
\includegraphics[
          height=4.9in,
         width=5.9in]{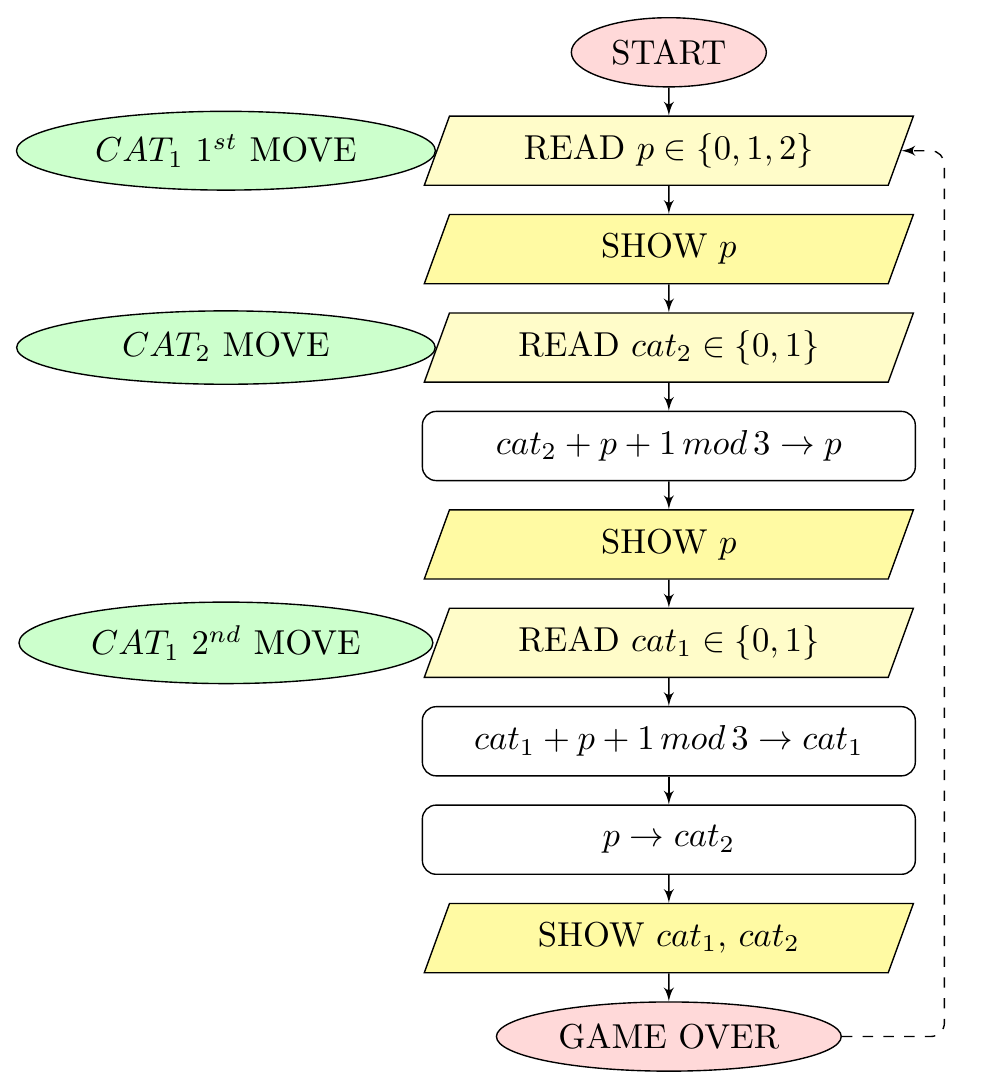}
			\caption{Flow diagram of each iteration of of the game.}\label{grg}
\end{figure}

The two players(\textit{Cat 1} i \textit{Cat 2})  are offered three foods: food number 0, food number 1, food number 2, always in pairs of two, in accordance with the following procedure. The first player (\textit{Cat 1}) selects and keeps one of the three foods. The second player (\textit{Cat 2}) chooses between the two remaining foods. The food kept in the first move and the food rejected by Cat 2 now form a pair from which a selection is made by \textit{Cat 1} ($2^{nd}$ move) . Diagram of the game is shown in Figure \ref{grg}.

The optimal behaviour of the players is not to favour any of the three foods (each being equally attractive and important). Both participants try to achieve a balanced diet. This assumption allows to extract meaning of intransitive orders,  because all the elements that we sort of are equally important.

\subsection{Mathematical description}

The first move of \textsl{Cat 1} is limited to select one of the three food. This movement can be described by the point $(P_0,P_1,P_2)$ of a three dimensional simplex, where $P_j$ denotes the frequency of the choice of food numbered by \textit{j}. Let us denote by $P(C_{k} | B_{j})$ the probability of choosing (by the Cat 2) food number \textit{k} when the offered pair of dishes does not contain food number \textit{j}. Analogously, number $Q(C_{k} | B_{j})$ denotes the probability of choosing by the \textit{Cat 1} food
number \textit{k} when the offered pair of dishes does not contain food number \textit{j}.\footnote{We refer here to the players mixed strategies, as pure strategies do not lead to optimal results.} Six conditional probabilities $P(C_{k} | B_{j})$ (or $Q(C_{k} | B_{j})$)  determine the behaviour of the players with respect to the proposed them in pairs food portions.

Let $\lambda_k$ and  $\omega_k$ denot the frequencies of appearance of the particular foods
in the \textit{Cat 1} and the \textsl{Cat 2} diet, respectively.

For  \textit{Cat 2} we obtain:
\begin{eqnarray}\label{starszy}
\omega_0&=&P(C_{0} | B_{1})P_1+P(C_{0} | B_{2})P_2,\nonumber\\
\omega_1&=&P(C_{1} | B_{0})P_0+P(C_{1} | B_{2})P_2,\\
\omega_2&=&P(C_{2} | B_{0})P_0+P(C_{2} | B_{1})P_1.\nonumber
\end{eqnarray}
From the perspective of the \textit{Cat 1}:
\begin{eqnarray}\label{mlodszy}
\lambda_0&=&Q(C_{0} | B_{1})[P(C_{1} | B_{0})P_0+P(C_{1} | B_{2})P_2]+Q(C_{0} | B_{2})[P(C_{2} | B_{0})P_0+P(C_{2} | B_{1})P_1],\nonumber\\
\lambda_1&=&Q(C_{1} | B_{2})[P(C_{2} | B_{0})P_0+P(C_{2} | B_{1})P_1]+Q(C_{1} | B_{0})[P(C_{0} | B_{1})P_1+P(C_{0} | B_{2})P_2],\\
\lambda_2&=&Q(C_{2} | B_{1})[P(C_{1} | B_{0})P_0+P(C_{1} | B_{2})P_2]+Q(C_{2} | B_{0})[P(C_{0} | B_{1})P_1+P(C_{0} | B_{2})P_2].\nonumber
\end{eqnarray}
Let us introduce the parameterization of the conditional probabilities.  

Let us assume:
\begin{eqnarray}\label{param}
P(C_2|B_0)=\frac{1+l_0}{2},&&  P(C_1|B_0)=\frac{1-l_0}{2},\nonumber \\
P(C_0|B_1)=\frac{1+l_1}{2},&&  P(C_2|B_1)=\frac{1-l_1}{2}, \\
P(C_1|B_2)=\frac{1+l_2}{2},&&  P(C_0|B_2)=\frac{1-l_2}{2}\,.\nonumber  
	\end{eqnarray}
and
	\begin{eqnarray}\label{param1}
Q(C_2|B_0)=\frac{1-L_0}{2},&&  Q(C_1|B_0)=\frac{1+L_0}{2},\nonumber \\
Q(C_0|B_1)=\frac{1-L_1}{2},&&  Q(C_2|B_1)=\frac{1+L_1}{2}, \\
Q(C_1|B_2)=\frac{1-L_2}{2},&&  Q(C_0|B_2)=\frac{1+L_2}{2}\,.\nonumber  
	\end{eqnarray}
\subsection{Optimal strategies}

The optimal strategy for the two players is to get the equal distribution of frequency of the occurrence  of a particular food in diet.
It is required that the following conditions are held:
\begin{eqnarray}\label{mlo}
\lambda_0=\lambda_1=\lambda_2&=&\frac{1}{3}\,,\\\label{str}
\omega_0=\omega_1=\omega_2&=&\frac{1}{3}\,. 
\end{eqnarray}
The (\ref{str}) condition takes the form:

\begin{eqnarray}\label{op}
l_1 P_1-l_2 P_2=\frac{2}{3}-(P_1+P_2),\nonumber \\
l_2 P_2-l_0 P_0=\frac{2}{3}-(P_0+P_2), \\
l_1 P_1-l_3 P_3=\frac{2}{3}-(P_0+P_3)\, ,\nonumber  
	\end{eqnarray}
and its solution:
\begin{eqnarray}\label{opr}
P_0=\frac{-(-1 - l_1 + l_2 - 3\,l_1 l_2)}{3 (1 + l_0 l_1 + l_0 l_2 + l_1 l_2)},\nonumber \\
P_1=\frac{-(-1 + l_0 - l_2 - 3\,l_0 l_2)}{3 (1 + l_0 l_1 + l_0 l_2 + l_1 l_2)}, \\
P_2= \frac{-(-1 - l_0 + l_1 - 3\,l_0 l_1)}{3 (1 + l_0 l_1 + l_0 l_2 + l_1 l_2)}\, ,\nonumber  
	\end{eqnarray}
defines a mapping of the three-dimensional cube $[-1,1]^3$ in the space of parameters $l_j$ (that define conditional probabilities $P(C_{k} | B_{j})$)  into a triangle--two-dimensional simplex. The
barycentric coordinates of a point of this triangle are interpreted as the probabilities $(P_0,P_1,P_2)$.

Note that \textit{Cat 2} is able to achieve the optimum effect (\ref{str}) if for every $j$ ($j=0,1,2$) the condition $P_j\leq\tfrac{2}{3}$ is satisfied. Indeed, if $P_0>\tfrac{2}{3}$ then $P_1+P_2<\tfrac{1}{3}$ and $\omega_0 <\tfrac{1}{3}$ (similarly in other cases).
Let us now consider the situation of (\textit{Cat 1}). This player moves first in each iteration of the game. The (\ref{mlo}) condition can be written as follows:
\begin{eqnarray}\label{stop}
(1 - L_1) (1 - l_0) P_0 + (1 + L_2) (1 + l_0) P_0 + (1 + L_2) (1 - l_1) P_1 + (1 - L_1) (1 + l_2) P_2 = 4/3,\nonumber \\
(1 - L_2) (1 + l_0) P_0 + (1 + L_0) (1 + l_1) P_1 + (1 - L_2) (1 - l_1) P_1 + (1 + L_0) (1 - l_2) P_2 = 4/3, \\
(1 + L_1) (1 - l_0) P_0 + (1 - L_0) (1 + l_1) P_1 + (1 - L_0) (1 - l_2) P_2 + (1 + L_1) (1 + l_2) P_2 = 4/3\, ,\nonumber  
	\end{eqnarray}
Let us suppose there are exist \footnote{If $P_j \leq\tfrac{2}{3}$ for any $j$.} parameters $l_i$ such that condition (\ref{str}) is satisfied -- Cat 2 reaches  the diet completeness. 
 
Substituting formula (\ref{opr}) on the probability $P_j$  into formula (\ref{stop}) immediately gives
\begin{eqnarray}\label{optw}
L_1 = L_2,\nonumber\\ L_0 = L_2, \\\nonumber L_0 = L_1.
\end{eqnarray}
These conditions describe a set of \textit{Cat's 1} optimal strategy.  This player must use strategy characterized by three (independent) equal conditional probabilities.  

In the next section we analyze the optimal strategies for both players in terms of their division into two types - intransitive and transitive.


\section{Optimal intransitive and transitive strategies}

We say that a player prefers food no. 1 to food no. 0 ($1\succ 0$) when he/she is willing to choose food no. 1 more often than food no. 0 from the offered pair $(0,1)$ ($P(C_1|B_2)>P(C_0|B_2)$). The situation corresponds to an intransitive choice if one of the following two conditions is satisfied:

	\vspace{0.1cm}
\begin{enumerate}
 \item $P(C_0|B_2)<P(C_1|B_2)$, $P(C_1|B_0)<P(C_2|B_0)$, $P(C_2|B_1)<P(C_0|B_1)$\,,
 \vspace{0.3cm}
 \item $P(C_0|B_2)>P(C_1|B_2)$, $P(C_1|B_0)>P(C_2|B_0)$, $P(C_2|B_1)>P(C_0|B_1)$\,. 
\end{enumerate}
These two conditions can be written (by using normalization probability measure to 1) in the following form:
\begin{eqnarray}\label{one}
 P(C_0|B_2)<\frac{1}{2}, P(C_1|B_0)<\frac{1}{2}, P(C_2|B_1)<\frac{1}{2}\,,\\
 P(C_0|B_2)>\frac{1}{2}, P(C_1|B_0)>\frac{1}{2}, P(C_2|B_1)>\frac{1}{2}\,. \label{two}
\end{eqnarray}

\begin{figure}[h]
\includegraphics[
          height=1.9in,
         width=2.1in]{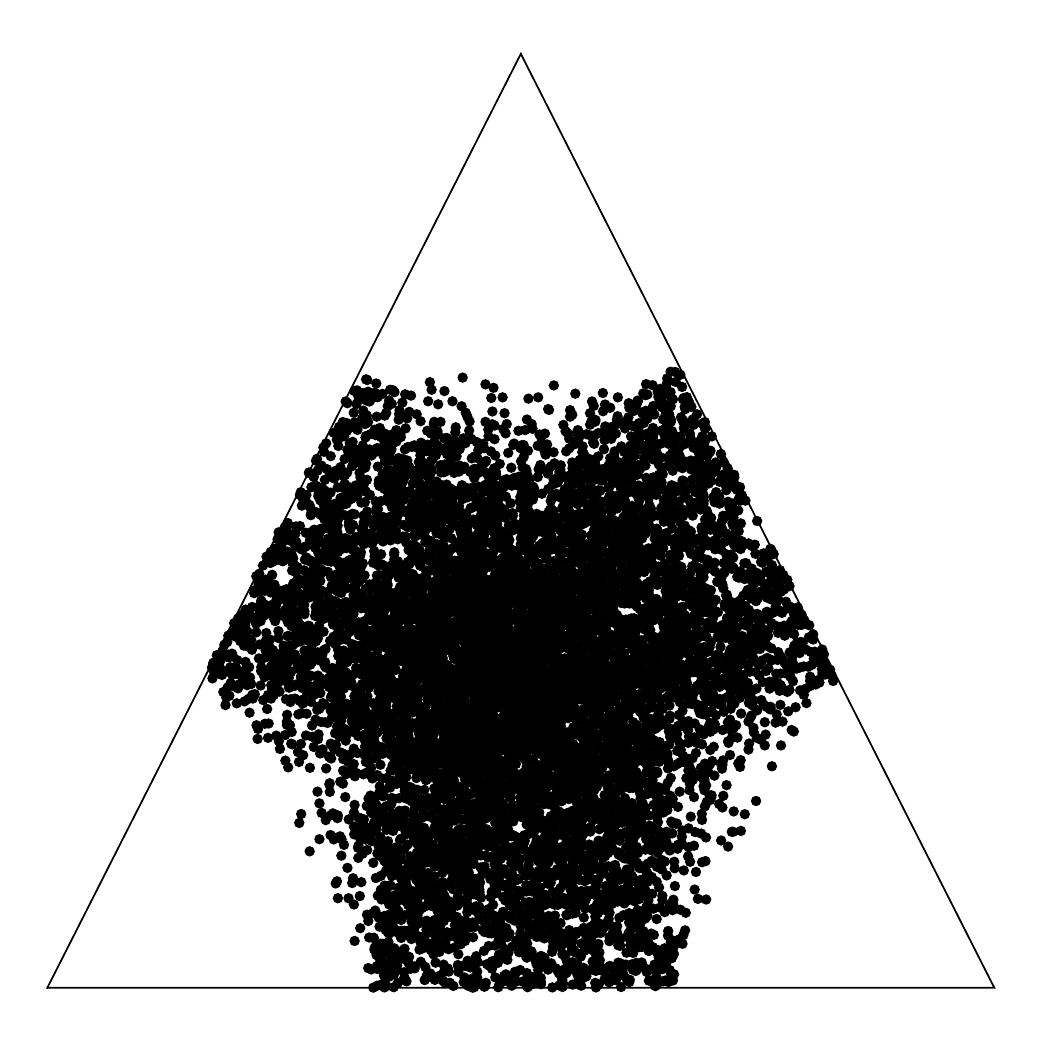}\includegraphics[
                      height=1.9in,
                      width=2.1
                     in]%
       {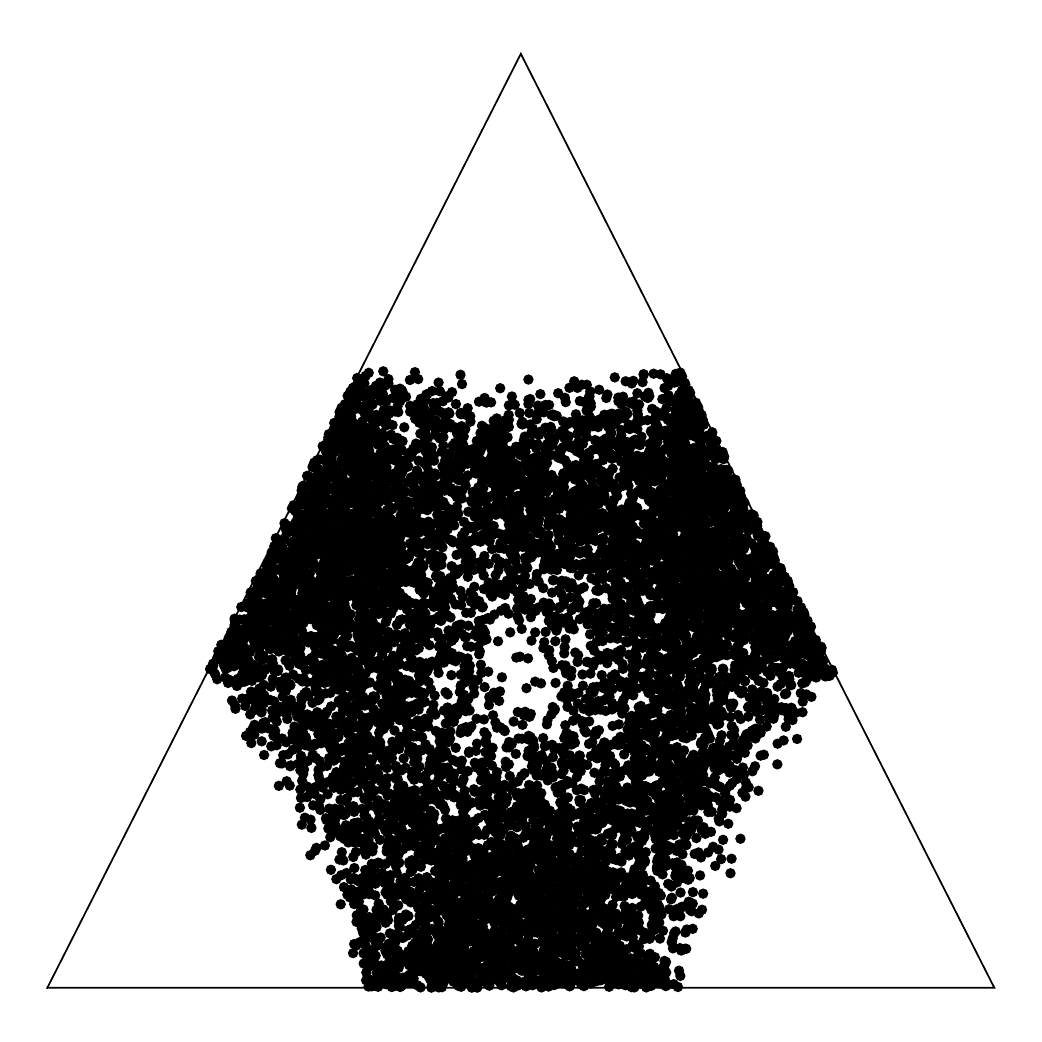}\includegraphics[
                      height=1.9in,
                      width=2.1
                     in]%
       {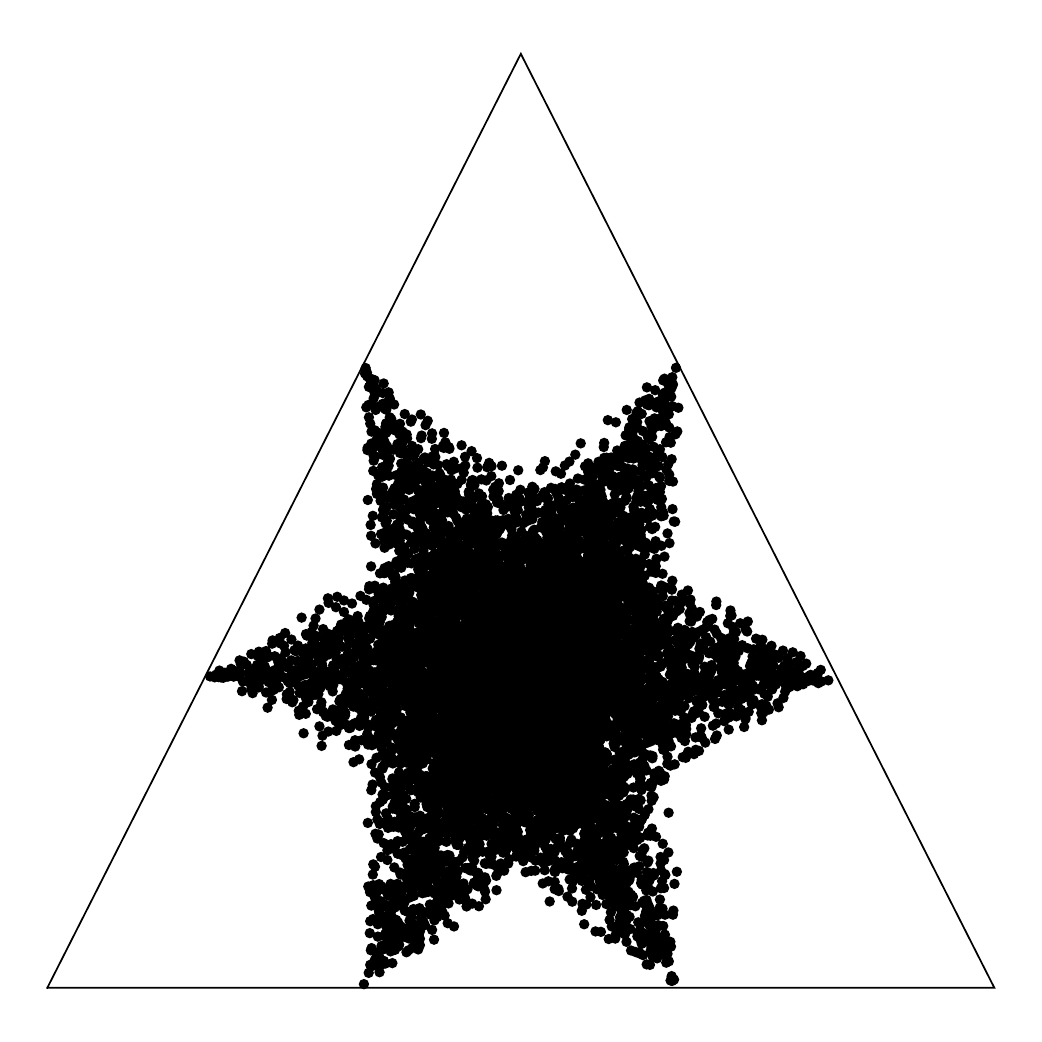}
			\caption{Availability (by \textit{Cat 2}) of different types of strategies. From left to right: all, transitive, intransitive.}
			\label{class}
\end{figure}

Let's see how different types of strategies of individual players are achievable.
The conditions under which \textit{Cat 2} makes his decision are similar to those considered in the work \cite{r1} (then it was a game with Nature). This player can select its optimal strategy if only $P_j\leq \tfrac{2}{3}$, for $j=0,1,2$. It turns out that under this assumption, this strategy can always be transitive, but not always intransitive. This can be illustrated graphically.
Figure \ref{class} presents the areas of frequency $(P_0,P_1,P_2)$ for which  \textit{Cat 2} optimal strategies exist. It is the range (for 10,000 randomly selected points)  of mapping (defined by equations (\ref{opr}) ) of the three-dimensional cube of parameters $t_j$ into a triangle $(P_0,P_1,P_2)$.

Probabilities $ (P_0, P_1, P_2) $, for which \textit{Cat 2} can identify the optimal intransitive strategy, form a six-armed star composed of two triangles (any of them corresponding to one of two possible intransitive orders).
Both the optimal strategies of any type and optimal transitive strategies correspond to regular hexagon specified by conditions
$P_j\leq \tfrac{2}{3}$, $j=0,1,2$. 

The  other player (\textit{Cat 1}) situation is completely different. \textit{Cat 1} selects at the final stage of each iteration of the game. This movement reveals his preference for one food over another from pair. 

As pointed out above, \textit{Cat 1} optimal strategies (assuming that the second player reaches the optimal strategy) satisfy the condition (\ref{optw}), therefore:
\begin{eqnarray}
Q(C_0|B_2)=Q(C_1|B_0)=Q(C_2|B_1).
\end{eqnarray}
This means that either (\ref{one}) or (\ref{two}) is satisfied. It means that \textit{Cat 1} has to make intransitive choices in
order to achieve the optimal result!

\section{Quantum cats}

\subsection{One-qubit pure strategies} 

In this section we use different method of obtaining conditional probabilities (provided by the quantum games theory) which describe the preferences of decision makers,\textit{Cat 1} and \textit{Cat 2}, over pairs of food. This method was first used in the work \cite{r2} and is based on the concept of the so-called mutually unbiased:

\begin{df}Two orthonormal bases $\mathcal{A}\equiv \{ \,|\psi_0 \rangle\negthinspace\,\ldots\,|\psi_{N-1} \rangle\negthinspace\,\}$  and~$\mathcal{B}\equiv \{ \,|\varphi_0 \rangle\negthinspace\,\ldots\,|\varphi_{N-1} \rangle\negthinspace\,\}$ in Hilbert space $ \mathbb{C}^N$ are mutually unbiased if
\begin{displaymath}
|\langle \psi_i | \varphi_j \rangle|^2 = \frac{1}{N},   
\end{displaymath}
for any $0\le i,j \le N-1$.
\end{df}
For two-dimensional Hilbert space, three mutually unbiased bases is given as follows
\begin{align}\label{bases}
 &\left\{ | 0 \rangle,| 1 \rangle \right\},&\nonumber \\ 
 &\left\{ \frac{| 0 \rangle+| 1 \rangle}{\sqrt{2}},\frac{| 0 \rangle-| 1 \rangle}{\sqrt{2}} \right\},&  \\ 
 &\left\{ \frac{| 0 \rangle+i | 1 \rangle}{\sqrt{2}},\frac{| 0 \rangle-i| 1 \rangle}{\sqrt{2}} \right\}.& \nonumber 
\end{align}
This set is the most important for our considerations. It is worth to mention here that mutually unbiased bases led Wiesner \cite{r27} to begin research into quantum cryptography, before asymmetric key cryptography was invented! These bases
play also an important role in universality of quantum market games \cite{r11,r12,28}.

Let us turn to the construction of conditional probabilities that define the strategy of the players.
Let us denote three different mutually unbiased bases of two-dimensional Hilbert space as
\begin{displaymath}
\{| 1 \rangle\negthinspace_{\,0}, | 2 \rangle\negthinspace_{\,0}\},
\{| 0 \rangle\negthinspace_{\,1}, | 2 \rangle\negthinspace_{\,1}\},
\{| 0 \rangle\negthinspace_{\,2}, | 1 \rangle\negthinspace_{\,2}\}=\{
(1,0)^{T},(0,1)^{T}\} 
\end{displaymath}
Strategy of choosing the food number \textit{k}, when the offered food pair not contain the food of number \textsl{l} is denoted by
 $| k \rangle\negthinspace_{\,l}$ ($k$,\,$l=0,1,2$ i~$k\ne l$).

 A family $\{|z\rangle\negthinspace\}$ ($z \in \overline{\mathbb{C}}$) of convex vectors:
 \begin{displaymath}
 | z \rangle\negthinspace:=| 0 \rangle\negthinspace_{\,2}+z
|1\rangle\negthinspace_{\,2}=| 0 \rangle\negthinspace_{\,1}+\frac{1-z}{1+z}
|2\rangle\negthinspace_{\,1}=| 1 \rangle\negthinspace_{\,0}+\frac{1+iz}{1-iz}
|2\rangle\negthinspace_{\,0},
\end{displaymath}
defined by the parameters of the heterogeneous coordinates of the projective
space $\mathbb{C}P^{1}$, represents all strategies spanned by the base vectors.

\begin{figure}[h]
\includegraphics[
          height=4.9in,
         width=5.9in]{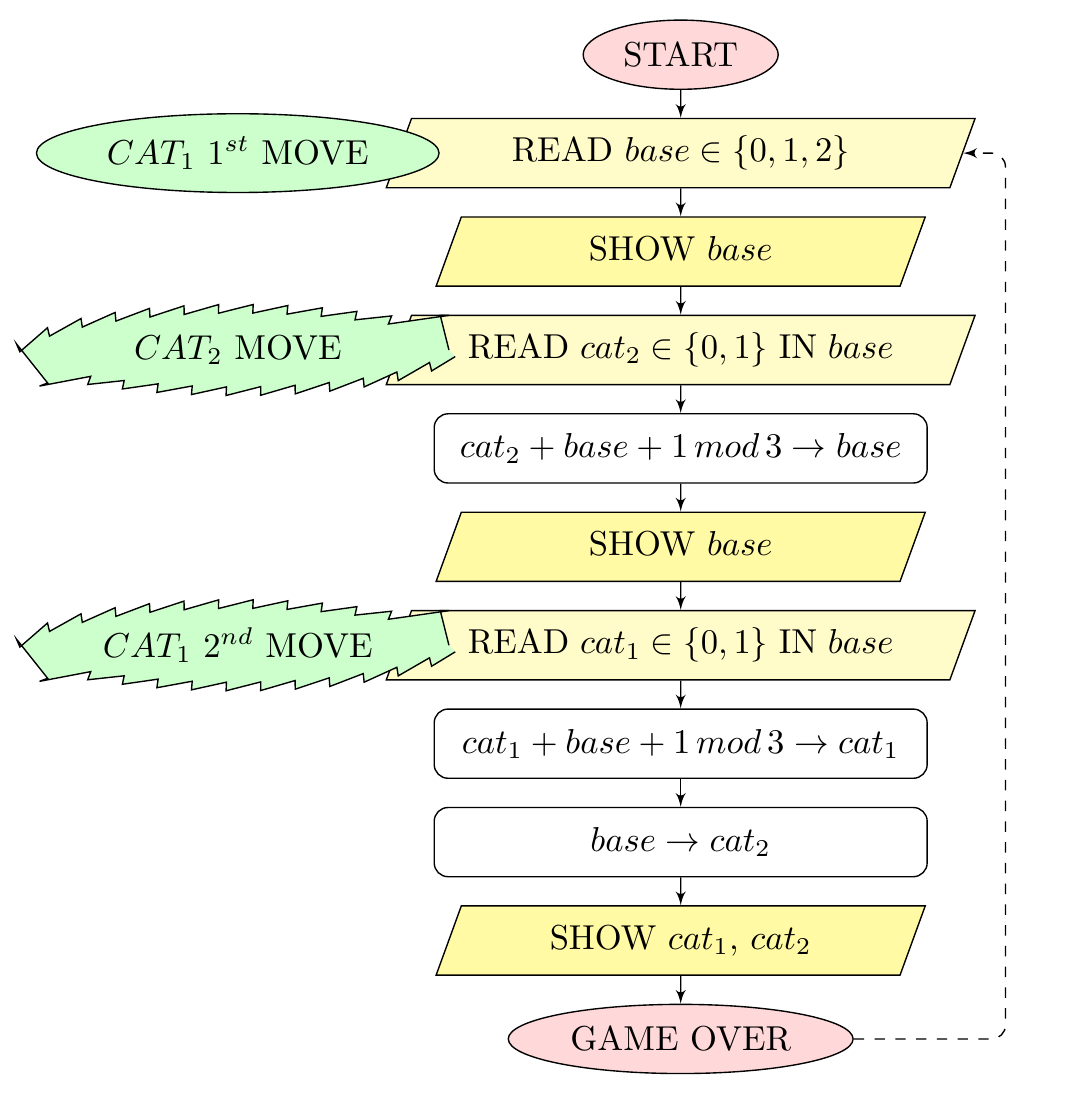}
			\caption{Flow diagram of each iteration of the quantum variant of game.}\label{grgo}
\end{figure}

The coordinates of the same strategy $|z\rangle\negthinspace$ read (measured) in three different bases (\ref{bases})
define quantum cats preferences toward a food pair represented by the base vectors.
Squares of their moduli, after normalization, measure the conditional probability of quantum cat’s making decision in choosing a particular product, when the choice is related to the suggested food pair. Diagram of the quantum variant of game is shown in Figure \ref{grgo}. It is similar to the previous one \ref{grg}, since the difference lies only in the process of implementing the game.

The player (\textit{Cat 1} or \textit{Cat 2} in our game) makes a decision to choose the right food from pair with the following probabilities:

\begin{align}\label{derek} 
 P(C_0|B_2) & =\frac{1}{1+|z|^{2}},&  P(C_1|B_2) &=\frac{|z|^{2}}{1+|z|^{2}}, \nonumber  \\
P(C_0|B_1)&=\frac{1}{1+|\frac{1-z}{1+z}|^2},
& P(C_2|B_1) &= 
\frac{|\frac{1-z}{1+z}|^2}{1+|\frac{1-z}{1+z}|^2}, \\
P(C_1|B_0)&=\frac{1}{1+|\frac{1+iz}{1-iz}|^2},&
P(C_2|B_0)&= \frac{|\frac{1+iz}{1-iz}|^2}{1+|\frac{1+iz}{1-iz}|^2}\,.\nonumber
\end{align}
It is convenient to parameterized $|z\rangle\negthinspace$ by the sphere $S_2\backsimeq \overline{\mathbb{C}}$ by using stereographic
projection which establishes bijection between elements of $\overline{\mathbb{C}}$ and the points of $S_2$.

The conditional probabilities can now be written in the following form:
\begin{align}\label{sphere}
P(C_0|B_2)&=\frac{1-x_3}{2},&  P(C_1|B_2)&=\frac{1+x_3}{2},&\nonumber \\
P(C_0|B_1)&=\frac{1+x_1}{2},&  P(C_2|B_1)&=\frac{1-x_1}{2},& \\
P(C_1|B_0)&=\frac{1+x_2}{2},&  P(C_2|B_0)&=\frac{1-x_2}{2}.&\nonumber  
	\end{align}
	
	Note that the probability (\ref{sphere}) are parameterized similarly to the classic model.
But in this case we have a sphere points, so the condition $x_1 ^ 2 + x_2^ 2 + x_3^ 2 = 1 $ must be fulfilled.
 
A careful reader certainly noticed that we abstained from introducing to much of the game-theoretical terminology. The problem can be easily reformulated in the language of quantum game theory \cite{rA6, rP7}. Another interesting approach would be to adopt the framework developed by A.~Lambert-Mogiliansky, I.~Martinez-Martinez \cite{29}, c.f.~also \cite{30}. This framework allows one to incorporate type indeterminacy of agents \cite{29,31,32}. In our model the agents are
of definite type.
\subsection{Availability of different types of one-qubit strategies}

\begin{figure}[h]
\includegraphics[
          height=1.9in,
         width=2.1in]{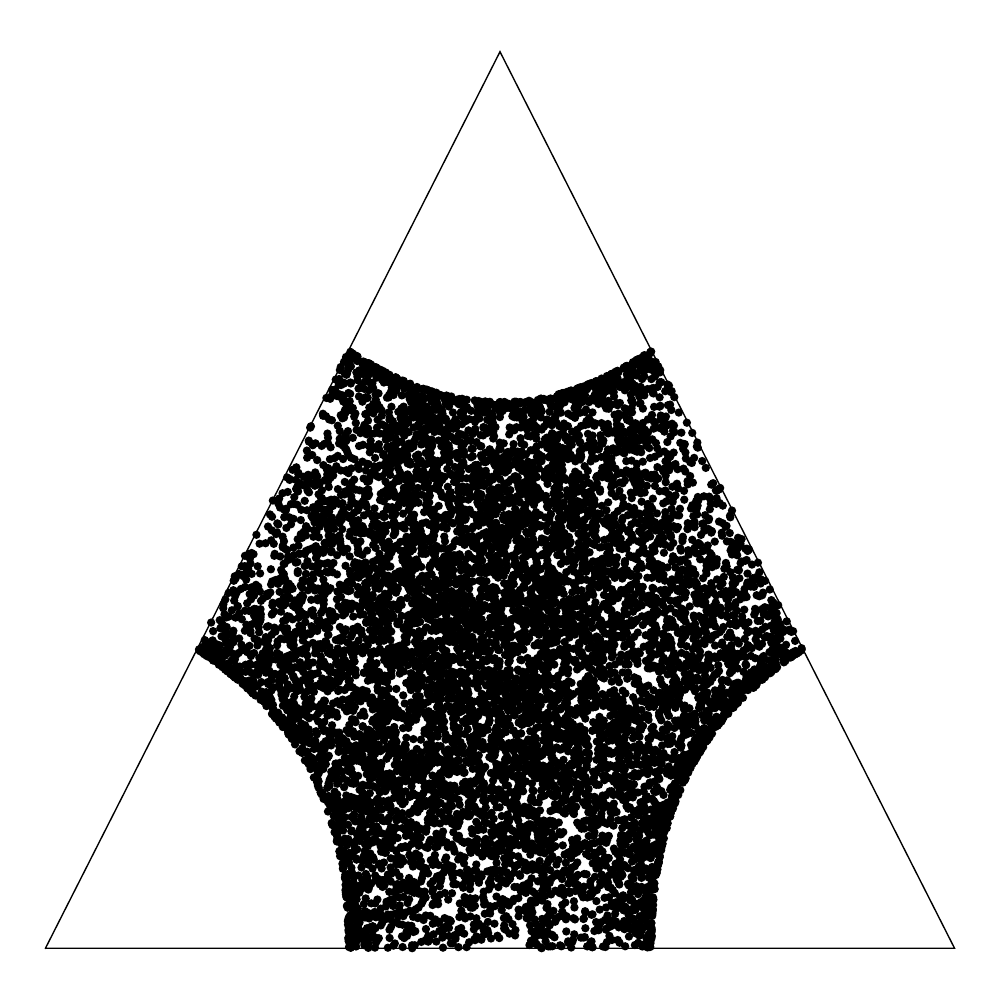}\includegraphics[
                      height=1.9in,
                      width=2.1
                     in]%
       {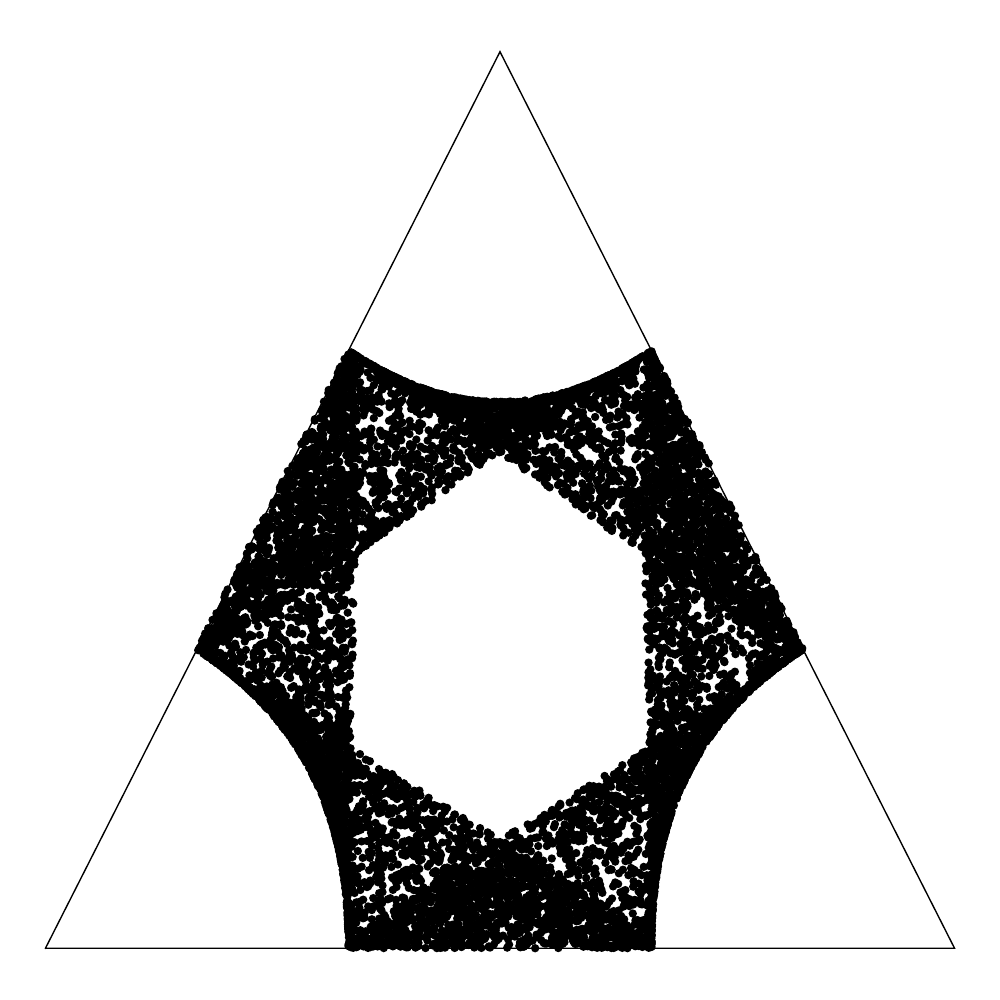}\includegraphics[
                      height=1.9in,
                      width=2.1
                     in]%
       {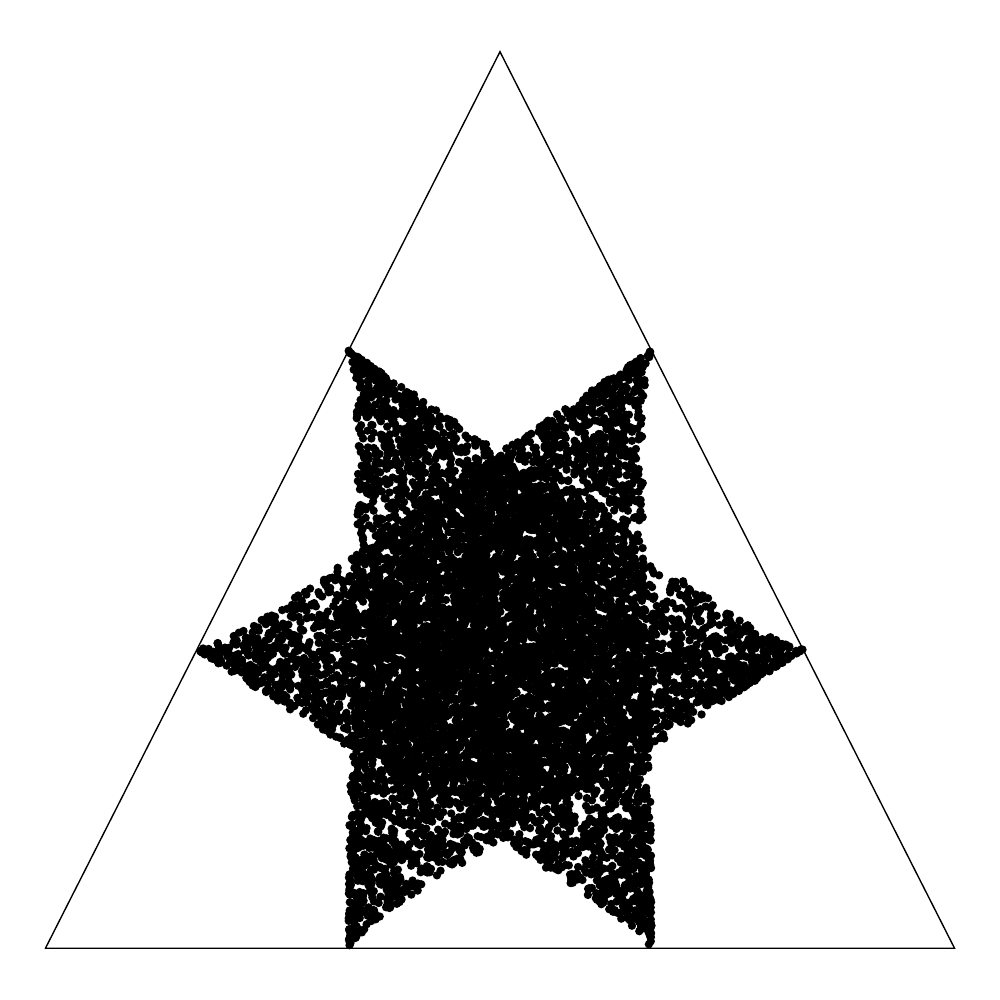}
			\caption{Availability (by \textit{Cat 2}) of different types of one-qubit strategies. From left to right: all, transitive, intransitive.}
			\label{qlass}
\end{figure}

Availability of various types of \textit{Cat 2}'s one-qubit optimal strategies  is illustrated in Figure \ref{qlass} (see \cite{r2}). 
Here we see a fundamental contrast between quantum and classical model. First of all, the area corresponding to the optimal transitive strategies does not cover the entire area of the optimal strategy of any type (transitiv or intransitiv). Transitive optimal strategies do not appear within the boundaries of hexagon-like figure in the central part of the triangle.

This means that for some frequencies $(P_0,P_1,P_2)$  \textit{Cat 2} is able to achieve optimal results only thanks to the intransitive strategy!
\footnote{Detailed quantitative analysis of second player's optimal strategies (Fig. \ref{class} and \ref{qlass}) is similar to that in work \cite{r2}. Readers interested in this, please refer to this paper.}

Let's analyze the \textit{Cat 1} situation in the case of one-qubit pure strategy, at the final stage of each iteration of the game. 
The probabilities $Q (C_k | B_j) $ are in the form (\ref{sphere}). To distinguish it from the strategy of the other player instead of $x_i$, we will use $ X_i $.

Assuming the \textit{Cat 2} achieves the optimal strategy, the optimality conditions (\ref{op}) for \textit{Cat 1} will simplify to:
\begin{eqnarray}\label{optwq}
X_1-X_3=0,\nonumber\\ X_2 +X_3=0, \\\nonumber X_1+X_2=0.
\end{eqnarray}
Since $ X_1^ 2 + X_2^2 + X_3^2 = 1 $ we obtain two optimal strategies that are characterized by two points $(X_1, X_2, X_3)$ 
of two-dimensional sphere:
\begin{eqnarray}\label{optwqP}
X_1= -\frac{1}{\sqrt{3}}, X_2=\frac{1}{\sqrt{3}}, X_3=-\frac{1}{\sqrt{3}},\nonumber\\\nonumber X_1= \frac{1}{\sqrt{3}}, X_2=-\frac{1}{\sqrt{3}}, X_3=\frac{1}{\sqrt{3}}.
\end{eqnarray}
Both of these strategies are intransitive. 

It is worth noting that in the language of quantum game theory one-qubit strategies are pure strategies, have an equal informative values (the zero entropy). Hence,  treating them in an equal way  is natural and does not raise any controversy similar of those of the constant measure (the Laplace's principle of insufficient reason) in the classical model. In this case, we used a mixed strategies which may provide different pieces of information. 
Pure strategies that we can identify with eight choice functions:
\begin{displaymath}
f_k:\{(0,1), (0,2),(1,2)\}\to \{0,1,2\},\qquad k=0,\ldots,7,
\end{displaymath}
do not lead us to identify the optimal strategy.


\section{Discussion and conclusions}

The variant of the game considered in the paper allows to explore the availability of different types of optimal strategies of both players (as opposed to earlier models which related to the situation of one player only). They make their decisions in a variety of contexts (depending on the decision made by the other player), which affects their possibility to use optimal intransitive and transitive strategies. The most interesting variant of the game has been analysed, i.e. such variant where the circumstances provide both players with the opportunity to achieve their optimal strategies.

It appears that the only optimal strategies of \textit{Cat 1} are intransitive strategies. This is indeed an interesting result. This player performs two moves in the game. In the first one, he chooses (with frequency $P_j$) one of the foods. The food is included in the pair from which he makes the final selection. Therefore, he has a partial influence on the composition of this pair\footnote{\textit{Cat 2} does not have this possibility .} 
It might seem that such a possibility should help him achieve an optimal strategy in the sense that he has more freedom in choosing the type of strategy. However, it is quite the opposite; he can only use an intransitive strategy. The other player (\textit{Cat 2}) in the classical variant of the game can always use a transitive strategy (in certain specific conditions also intransitive strategy, see Fig. \ref{class}).

Replacing classical strategies of the players with one-qubit quantum strategies will increase the importance of intransitive strategies. At certain $(P_0,P_1,P_2)$ frequencies, in order to achieve an optimal effect, both players must apply intransitive strategies! Moreover, the one-qubit strategies are pure strategies -- have an equal mathematical and information status (in contrast to the mixed strategies in classical model).

The model discussed in the paper, despite its mathematical simplicity, clearly illustrates how the type of preferences depends on the context of the decision to be made and on the behaviour of other players. The rejection of intransitive strategies as undesirable strategies is an unnecessary limitation. A distinction between intransitive and transitive strategies does not necessarily mean a distinction between irrational and rational strategies. However, it may well describe the decision-making process from an entirely different point of view -- the context of choice (e.g. whether and how the freedom of choice with regard to the decisions we take is affected by the adopted rules or the behaviour of other players). 

 It is worth mentioning that at present we can observe rapidly growing literature where ideas of Quantum Mechanics are proposed to explain a problems of behavioral and social science \cite{33,34,35,36,37,38}. The formalism of quantum mechanics can be useful in providing explanation to violations of transitivity in decision-making process (see \cite{30}).  We are still not fully aware of the implications of quantum theory in computer science models, machine learning and especially in decision-making. The work that has been done in this field indicates significant progress in the development this part of our knowledge.    












\section*{Acknowledgments}

This work was supported by the Polish National Science Centre under the project
number \textbf{DEC-2011/01/B/ST6/07197}.









\begin{thebibliography}{999} 


\bibitem{ecbi} von Neumann, J.; Morgenstern, O. {\em Theory of Games and Economic Behavior}.  Princeton University Press, Princeton, US, 1944.
\bibitem{stt}  Straffin, P. D. {\em Game Theory and Strategy}. The Mathematical Association of America, Washington DC, US, 1992. 

\bibitem{inf} Rasmusen, E. {\em Games and Information: An Introduction to Game Theory}. Wiley-Blackwell, Malden, US, 2007.

\bibitem{rq} Nielsen, M. A,; Chuang, I. L. {\em Quantum Computation and Quantum information}.  Cambridge University Press, New York, US, 2000.

\bibitem{re4} Eisert, J.; Wilkens, M.; Lewenstein, M.  Quantum games and quantum strategies. {\em Phys. Rev. Lett.} {\bf 1999}, {\em 83}, 3077-3080.

\bibitem{rd5} Meyer, D. Quantum strategies. {\em Phys. Rev. Lett.} {\bf 1999}, {\em 8}, 1052-1055.

\bibitem{rA6} Flitney, A. P.; Abbott, D.  An introduction to quantum game theory. {\em Fluct. and Noise Lett.} {\bf 2002}, {\em 2} R175-R187.

\bibitem{rP7} Piotrowski, E. W.; S\l{}adkowski, J.  An invitation to quantum game theory. {\em Inter. J. of Theor. Phys.} {\bf 2003}, {\em 42}, 1089-1099.

\bibitem{r8} Piotrowski, E. W.; S\l{}adkowski, J.  The next stage: quantum game theory, In {\em Mathematical Physics Research at the Cutting Edge}, C. V. Benton, C. V. Eds.; Nova Science Publishers: New York, US, 2004; pp. 247-268.

\bibitem{r1}  Piotrowski, E. W.;  Makowski, M.  Cat's dilemma - transitivity vs. intransitivity. {\em Fluct. and Noise Lett.} {\bf 2005}, {\em 5}, L85-L96.

\bibitem{r2}  Makowski, M.;  Piotrowski, E. W.  Quantum cat's dilemma: an example of intransitivity in a quantum game. {\em Phys. Lett. A} {\bf 2006}, {\em 355}, 250-254.
 
\bibitem{r2n} Makowski, M. Transitivity vs. intransitivity in decision making process - an example in quantum game theory {\em Phys. Lett. A} {\bf 2009}, {\em 373}, 2125-2130. 

\bibitem{En} Makowski, M.;  Piotrowski, E. W. Transitivity of an entangled choice.  {\em J. Phys. A Math. Theor.} {\bf 2011}, {\em 44}, {\em 075301}, 1-12.

\bibitem{El} Makowski, M.;  Piotrowski, E. W. Decisions in elections - transitive or intransitive quantum preferences. {\em J. Phys. A Math. Theor.} {\bf 2011}, {\em 44, 215303}, 1-10.

\bibitem{div} Makowski, M.;  Piotrowski, E. W.  When I cut, you choose method implies intransitivity. {\em Physica A} {\bf 2014}, {\em 415}, 189-193.

\bibitem{ST}  Steinhaus, H.  {\em Memoires and Notes} (in Polish), Aneks, London, UK, 1992.

\bibitem{trra} Anand, P.; Pattanaik, P. K.; Puppe, C. Introduction, in: {\em Handbook of Rational and Social Choice};  Anand, P.; Pattanaik, P. K. and Puppe, C. Eds.; Oxford University Press, US, 2009; pp.1--20.

\bibitem{wr}  Von Wright, G. H. The Logic of Preference: an Essay,  Edinburgh University Press, Edinburgh, UK, 1963.

\bibitem{r33} Grosenick, L.; Clement, T. S.;  Fernald, R. D. Fish can infer social rank by observation
alone. {\em Nature}, {\bf 2007} {\em 445}, 429-432.

\bibitem{tull} Tullock, G.  The irrationality of intransitivity. {\em Oxford Economic Papers}  {\bf 1964}, {\em 16}, 401--406.

\bibitem{r23} Fishburn, P. C.  Nontransitive Preferences in Decision Theory. {\em J. of Risk and Uncertainty} {\bf 1991} {\em 4(2)}, 113-134.

\bibitem{tw} Fishburn, P. C.  Nontransitive measurable utility. {\em J. of Math. Psychology} {\bf 1982}, {\em 26}, 31-67.
 

\bibitem{r24}  Arrow, K. J.  {\em Social Choice and Individual Values}, Yale Univ. Press, New York, US, 1951.

\bibitem{lek} Poddiakov, A. N. Intransitive Character of Superiority Relations and Decision-making. {\em Psychology. The J. of the Higher School of Econ.} {\bf 2006}, {\em Vol. 3 N 3},  88-111.

\bibitem{T} Tversky, A. Intransitivity of preferences. {\em Psychological Rev.} {\bf 1969}, {\em 76(1)}, 31-48.

\bibitem {Gar} Gardner, M.  Mathematical Games: The Paradox of the Nontransitive Dice and the Elusive Principle of Indifference. {\em Sci. Amer.} {\bf Dec. 1970}, {\em 223}, 110-114.

\bibitem{r35} Penney, W. Problem 95: penney-ante. {\em J. of Rec. Math.} {\bf Oct. 1969}, 241-258.

\bibitem{r15} Boddy, L.  Interspecific combative interactions between wood-decaying basidiomycetes. {\em FEMS Microbiol Ecol.} {\bf 2000}, {\em 31}, 185-194.

\bibitem{jasz} Sinervo, B.; Lively, C. M.  The rock-paper-scissors game and the evolution of alternative male strategies. {\em Nature} {\bf 1996}, {\em 380}, 240-243.

\bibitem{r16} Shafir, S. Intransitivity of preferences in honey bees: support for'comparative'evaluation of foraging options. {\em Animal Behaviour} {\bf 1994}, {\em 48}, 55-67.

\bibitem{ope} Pahikkala, T.;  Waegeman, W.; Tsivtsivadze, E.;  Salakoski, T.; De Baets, B. Learning intransitive reciprocal relations with kernel methods. {\em European J. Oper. Res.} {\bf 2010}, {\em 206(3)}, 676-685.

\bibitem{Ki}  Klimenko, A. Y. Abstract competition and competitive thermodynamics. {\em Phil. Trans. R. Soc. A} {\bf 2013}, {\em 371(1982)}, 1-16.

 \bibitem{H}  Halpern, J. Y.  Intransitivity and vagueness. {\em Rev. Symbol. Log.} {\bf 2008}, {\em 1(04)}, 530-547.

\bibitem{Trv} Tversky, A.; Kahneman, D. {\em Choices, Values, and Frames}. Cambridge Univ. Press, New York, US, 2000.

\bibitem{St} Karni, E. {\em Decision Making Under Uncertainty: The Case of State-Dependent Preference}. Harvard Univ. Press, Cambridge, UK, 1985.

\bibitem{trS} Tversky, A.; Simonson, I.  Context-dependent preferences. {\em Management Science} {\bf 1993}, {\em 39}, 1179-1189.

\bibitem{De} Aerts, D.; Sozzo, S.; Veloz, T. Quantum Structure in Cognition and the Foundations of Human Reasoning. {\bf 2014}, arXiv:1412.8704v1 [cs.AI], http://arxiv.org/pdf/1412.8704.pdf.

\bibitem{kier} Khrennikova, P,;  Haven, E.; Khrennikov, A.  An application of the theory of open quantum systems to model the dynamics of party governance in the US Political System. {\em Int. J. of Theor. Phys.} {\bf 2013}, {\em 53 (4)}, 1346-1360.

\bibitem{Brams}  Brams, S. J.; Taylor, A. D. {\em Fair Division - From Cake-cutting to Dispute Resolution}. Cambridge Univ. Press, New York, US, 1996.

\bibitem{ST1} Steinhaus, H. {\em Mathematical snapshots}.  Oxford Univ. Press, New York, US, 1996.



\bibitem{r27} Wiesner, S.  Conjugate coding. {\em Sigact News} {\bf 1983}, {\em 15}, 78-88.

\bibitem{r11} Piotrowski, E. W.; S\l{}adkowski, J.  Quantum Market Games. {\em Physica A} {\bf 2002}, {\em 312}, 208-216.

\bibitem{r12} Piotrowski, E. W.; S\l{}adkowski, J. Quantum-like Approach to Financial Risk: Quantum Anthropic Principle,  {\em Acta Physica Polonica B} {\bf 2001}, {\em 32}, 3873-3879.

\bibitem {28} Paku\l{}a, I., Piotrowski, E. W.; S\l{}adkowski, J. Universality of measurements on quantum markets.
{\em Physica A} {\bf 2007}, {\em 385}, 397-405.

\bibitem{29} Lambert-Mogiliansky, A.; Martinez-Martinez, I. Basic Framework for Games with Quantum-like Players. {\em PSE Working Papers} {\bf 2014}, {\em 2014-42}, 1-21.

\bibitem{30} Lambert-Mogiliansky, A.; Zamir, S.; Zwirn, H. Type indeterminacy: A model of the KT(Kahneman-Tversky)-man. {\em J. of Math. Psych.} {\bf 2009},  {\em 53 (5)}, 349-361.

\bibitem{31} Kvam, P. D.; Busemeyer, J. R.; Lambert-Mogiliansky, A. An Empirical Test of Type-Indeterminacy in the Prisoner's Dilemma. In {\em Quantum Interaction}; Atmanspacher, H., Haven, E., Kitto, K., Raine, D., Eds; Springer Berlin Heidelberg; 2014; pp 213-224.

\bibitem{32} Lambert-Mogiliansky, A.; Busemeyer, J. R.  Quantum type indeterminacy in dynamic decision-making: Self-control through identity management. {\em Games} {\bf 2012}, {\em 3(2)}, 97-118.

\bibitem{33} Brandenburger, A.,  La Mura, P. Team Decision Problems with Classical and Quantum Signals.  arXiv:1107.0237v3 [quant-ph] {\bf 2015}, 1-18.

\bibitem{34} Busemeyer, J. R.; Wang, Z.; Townsend, J. T. Quantum dynamics of human decision making. {\em J. of Math. Psych.} {\bf 2006}, {\em 50},  220-241.

\bibitem{35} Brandenburger, A. The relationship between quantum and classical correlation in games. {\em Games and Economic Behavior} {\bf 2010}, {\em 69}, 175-183.

\bibitem{36} La Mura, P. A Double-Slit Experiment for Non-Classical Interference Effects in Decision Making.
{\em Topics in Cognitive Science}, {\bf 2014}, {\em 6}, 58-62. 

\bibitem{37} Yukalov, V.; Sornette, D. Conditions for Quantum Interference in Cognitive Sciences. {\em Topics in Cognitive Science}, {\bf 2014}, {\em 6},  79-90.

\bibitem{38} Busemeyer, J. R.; Bruza, P. D. {\em Quantum Models of Cognition and Decision}. Cambridge Univ. Press, New York, US, 2012.

\end{thebibliography}


%


%

\end{document}